\documentstyle{rnote}                
\input epsf.tex                      

\begin{document}



\title
{Resonant Tunnelling through InAs Quantum Dots\\
in Tilted Magnetic Fields: Experimental Determination\\
of the g-factor Anisotropy}
\author{{\sc J.~M.~Meyer} (a), {\sc I.~Hapke-Wurst} (a),\\
{\sc U.~Zeitler} (a), {\sc R.~J.~Haug} (a),
and {\sc K.~Pierz} (b)
}
\address{(a) Institut f\"ur Festk\"orperphysik, Universit\"at Hannover,\\
Appelstra{\ss}e 2, 30167 Hannover, Germany\\
(b) Physikalisch-Technische Bundesanstalt Braunschweig,\\
Bundesallee 100, 38116 Braunschweig, Germany}
\submitted{July 14, 2000}
\maketitle

subject classification: 73.40 Gk, 73.23 Hk, 73.61 Ey; S7.12

\begin{abstract}
We have determined the Land\'e factor $g^*$ in self-organized InAs
quantum dots using resonant-tunnelling experiments. With the
magnetic field applied parallel to the growth direction $z$ we find 
$g^*_\parallel = 0.75$ for the specific dot investigated.  
When the magnetic field is tilted away by $\vartheta$ from the growth axis,
$g^*$ gradually increases up to a value $g^*_\perp = 0.92$ when
$B \perp z$. Its angular dependence is found
to follow the phenomenological behaviour 
$g^* (\vartheta) = \sqrt{ (g^*_\parallel \cos\vartheta)^2   
                         +(g^*_\perp \sin\vartheta)^2 }$.  

\end{abstract}



\section{Introduction}

Resonant tunnelling experiments through zero-dimensional structures
are an efficient tool to access their quantized energy 
levels. In recent years several groups succeeded in performing
such experiments with self-assembled InAs quantum dots (QDs) embedded
in the barrier of a single-barrier tunnelling device~\cite{tunnel,SST}.
Low-temperature experiments allowed to measure directly the Land\'e factor 
$g^*$ of InAs dots~\cite{Andy,wir}. In this Note we will report on 
resonant tunnelling experiments through InAs QDs in tilted magnetic fields.
We will show that $g^*$ depends on the 
orientation of the magnetic field with respect to the
growth direction and can be described phenomenologically
by a tensor with two independent components.

\section{Sample structure}

The samples were grown by molecular-beam epitaxy on a highly Si-doped 
GaAs substrate with a donor concentration $n=2\times 10^{18}$~cm$^{-3}$.
First  we grew a 1-$\mu$m thick GaAs buffer layer with the same 
doping level followed by two 10-nm thick n-doped GaAs layers with 
$n=1\times 10^{17}$~cm$^{-3}$ and $n=1\times 10^{16}$~cm$^{-3}$ 
and an 15-nm thick undoped
GaAs spacer. On this bottom electrode a 10-nm thick AlAs barrier
was deposed.
The growth of the barrier was interrupted at a thickness of 5 nm
where 1.8 mono-layers of InAs were embedded in the barrier.
With such an InAs coverage self-assembled InAs QDs are formed.
The structure was terminated with a top electrode symmetric
to the bottom electrode finishing with
1~$\mu$m highly n-doped GaAs ( $n=2\times 10^{18}$~cm$^{-3}$).
As a consequence
of the high doping three-dimensional electrodes are present on both
sides of the AlAs barrier.

Subsequent to the growth of the wafer macroscopic AuGeNi contacts 
with a typical diameter of 50~$\mu$m were annealed into the top
electrode and vertical tunnelling diodes with the same diameter
were processed using wet-chemical etching.

\section{Resonant Tunnelling and Spin Splitting}

\begin{figure}[t]
\begin{tindent}            
\epsfxsize=5cm             
\epsffile{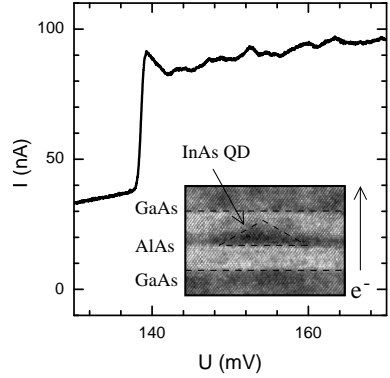} 
\end{tindent}              
\caption{( Small secCion of the $I$-$V$-curve 
of a single barrier GaAs-AlAs-GaAs tunnelling device
with InAs dots embedded in the middle of the AlAs barrier. 
The structure of a reference sample is
shown in the transmission electron micrograph 
in the inset. At positive bias
voltages the electrons tunnel from the bottom to the top as
indicated by the arrow. The step observed is due to 
single electron tunnelling through an individual InAs quantum dot.}
\label{step}
\end{figure}

When a bias voltage is applied between the top and the bottom electrode
our devices show $I$-$V$-characteristics which can be described 
by tunnelling through a single AlAs barrier~\cite{SST}.
Superimposed on this coarse $I$-$V$-curve we observe distinct current
steps assigned to resonant tunnelling through individual InAs quantum 
dots~\cite{tunnel}, for more details see~\cite{SST,wir}.
A typical step for $T=0.5$~K is shown in Fig.~\ref{step}.

\begin{figure}[t]
\begin{tindent}            
\epsfxsize=12cm             
\epsffile{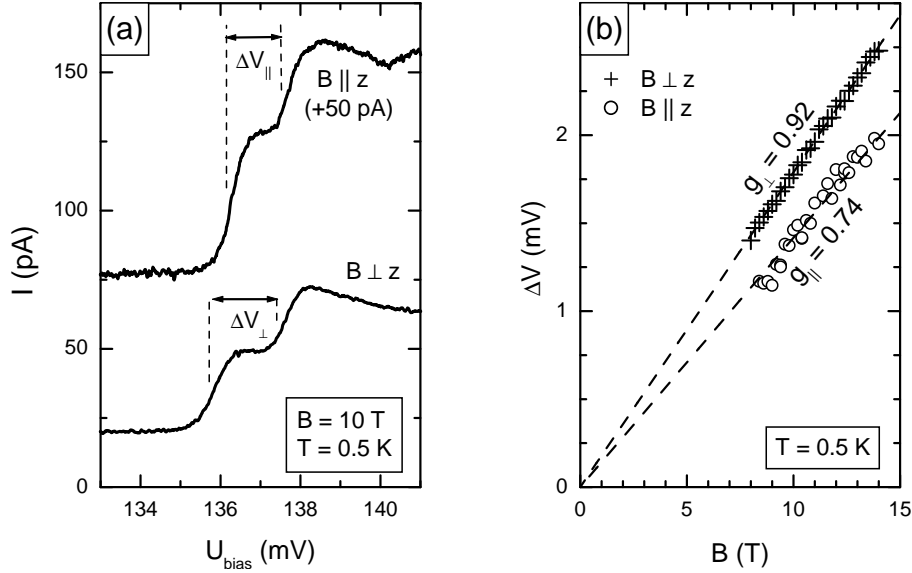}        
\end{tindent}              
\caption{(a) Spin splitting of the current step shown in Fig.~\ref{step} for
$B=10$~T at $T=0.5$~K for two field orientations.
The top trace is shifted for clarity.\newline
(b) Dependence of the spin splitting $\Delta V$ on the magnetic field
for both field orientations.}
\label{split}
\end{figure}

In high magnetic fields the
current step observed at $B=0$~T develops 
into two spin-split steps with a voltage separation
$\Delta V = g^* \mu_B B / \alpha e$. 
Here $\alpha=E_D/eV$
denotes the lever factor between the
energy separation emitter-dot, $E_D$, and the total
voltage drop, $V$. From the temperature dependent 
smearing of the current step we deduced $\alpha = 0.3$.
The spin splitting of the current step is shown in Fig.~\ref{split}a 
where the $I$-$V$-characteristic of the tunnelling structure
at $B=10$~T is displayed. For the top curve the magnetic field
$B$ is oriented along the growth direction $z$, the bottom trace
was measured in a magnetic field perpendicular to $z$.

As can be seen in Fig.~\ref{split}b, $\Delta V$ indeed increases
linearly with magnetic field confirming the scenario of
a simple Zeeman splitting of the QD's energy level.
However, the Land\'e factor $g^*$ as deduced from the
slope is considerably different for the two
field orientations $B \parallel z$ and $B \perp z$ where we 
find Land\'e factors $g^*_\parallel = 0.74$
and $g^*_\perp =0.92$~\renewcommand{\thefootnote}{\fnsymbol{footnote}}\footnote[2]{\sl
We obtained similar results 
for a second step observed at higher bias voltages.
This step can be assigned to the tunneling through
another InAs QD.
For the two field orientations we find $g^*_\parallel = 0.78$
and $g^*_\perp =0.99$.}.
From the temperature dependence of the step heights in
high magnetic fields we find that the low-voltage step
can be assigned to tunnelling of spin-down electrons\cite{wir}
resulting in a positive $g$-factor as also found
in~\cite{Andy}.


At first sight it is quite astonishing that the Land\'e factor is
far away from that of bulk InAs ($g^* = - 14.8$).
This  discrepancy can be explained qualitatively considering
effects of size quantization, 
strain and possible other effects~\cite{Andy}.
Such effects may be the leakage of the electronic
wave-function in the InAs quantum dot into the AlAs barrier
and the space dependent alloying of the InAs dots with AlAs.
In a simple picture this can be described by
an (rather complex) admixture of the Land\'e factor in AlAs
and strained InAlAs to the (size-quantized) $g$-factor in the InAs-dot.

\section{Tilted Magnetic Fields}

\begin{figure}[t]
\begin{tindent}            
\epsfxsize=5cm             
\epsffile{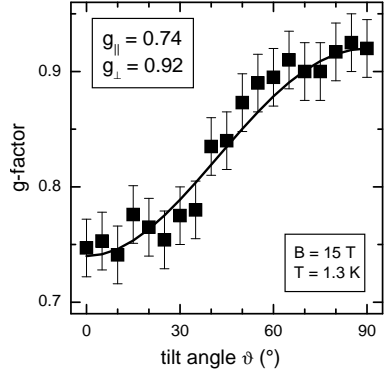} 
\end{tindent}              
\caption{Angular dependence of the experimentally measured $g^*$.
The dots show the experimental values, the solid line is a
phenomenological fit with two independent tensor components
of the $g$-factor tensor.}
\label{gvontheta}
\end{figure}

In order to clarify the dependence of $g^*$ on the orientation
of the magnetic field more clearly we have
performed experiments in tilted magnetic fields at a temperature
$T = 1.3$~K.
The angle between the magnetic field and the growth
direction was stepped from $\vartheta = 0^o$ to $\vartheta = 90^o$. 
For each angle I-V-curves were recorded at $B = 15$~T and the voltage
splitting between two spin-split steps was determined.
From this we deduced the angular dependent Land\'e factor, 
$g^*(\vartheta)$. The results are shown in Fig.~\ref{gvontheta}.
As can be seen in the figure, $g^* (\vartheta)$ gradually
increases when tilting the magnetic field away from the growth axis.
As indicated with the solid line it follows an angular dependence 
\begin{equation}
g^* (\vartheta) = \sqrt{ (g^*_\parallel \cos\vartheta)^2   
                         + (g^*_\perp \sin\vartheta)^2 },
\label{gtheta}
\end{equation}
with $g^*_\parallel=0.74$ and $g^*_\perp=0.92$.

This phenomenological behaviour can be understood when regarding the 
Zeeman contribution to the total energy in a size quantized structure.
For the most general case the corresponding Hamiltonian can be
written in the form~\cite{Kiselev}
\begin{equation}
{\cal H}_Z = \frac{1}{2} \mu_B \sigma_\alpha g_{\alpha \beta} B_\beta~,
\label{H}
\end{equation}
were $\alpha = x,y,z$ are the three spatial directions, $B_\alpha$ are
the components of the magnetic field along $\alpha$, $\mu_B$ is the
Bohr magneton and $\sigma_\alpha$ are the Pauli spin-matrices.
In the most general form the tensor $g_{\alpha\beta}$ contains nine
independent real components~\cite{Kiselev}.

To get a better physical insight we model our dot by a flat disc
with a height $h \approx 3$~nm and a diameter 
$d \approx 15$~nm~\cite{SST}.
In this case the Land\'e tensor reduces to a diagonal tensor with
two independent components $g^*_\perp = g_{xx} = g_{yy}$ and
$g^*_\parallel = g_{zz}$. In a tilted magnetic field with 
$B_x=0$, $B_y = B \sin\vartheta$ and $B_z = B \cos\vartheta$ 
equation~(\ref{gtheta}) then directly follows from (\ref{H}).

\section{Discussion}
Our experiment clearly shows the anisotropic nature 
of spin splitting in InAs quantum dots. 
Comparing with the measured $g$-factor anisotropy in 
quantum-wires~\cite{Oestreich} and quantum wells~\cite{LeJeune}
one would expect the largest $g$-factor when the magnetic
field is applied in the direction of the strongest confinement.
Our experiments, however, indicate that this simple relation for 
the $g$-factor anisotropy does not hold. We observe a smaller 
$g^*$  if the magnetic field is applied in the growth direction
where the quantum confinement is the strongest. Therefore, 
it is probable that the effects responsible 
for the $g$-factor anisotropies are not dominated by quantum
confinement but by more complicated mechanisms such as AlAs alloying
into the InAs quantum dots, leakage of the electronic wave function
into the AlAs barriers, spatial dependent strain, spin interactions 
between the dot and the electrodes etc. 
Without any doubt 
more elaborate theoretical models are necessary to clarify the 
possible origins of $g$-factor anisotropies in self-assembled
InAs quantum dots.

\section{Conclusions}
Using resonant tunnelling experiments we have measured the 
Land\'e factor tensor of InAs quantum dots. 
The $g$-factor tensor was described by two independent
tensor components $g^*_\parallel$ and $g^*_\perp$.








\begin{references}

\bibitem{tunnel}{\sc I.~E.~Itskevich, T.~Ihn, A.~Thornton, M.~Henini,
T.~J.~Foster, P.~Moriarty, A.~Nogaret, P.~H.~Beton, L.~Eaves}, 
and {\sc P.~C.~Main}, 
Phys.\ Rev.\ B\ {\bf 54}, 16401 (1996);\\      
{\sc T.~Suzuki, K.~Nomoto, K.~Taira}, and {\sc I. Hase},
Jpn.\ J.\ Appl.\ Phys.\ {\bf 36}, 1917 (1997);\\
{\sc M.~Narihiro, G.~Yusa, Y.~Nakamura, T.~Noda}, and {\sc H.~Sakaki},
Appl.\ Phys.\ Lett.\ {\bf 70}, 105 (1997).



\bibitem{SST}
 {\sc I.~Hapke-Wurst, U.~Zeitler, H.~W.~Schumacher, R.~J.~Haug, K.~Pierz}, 
and {\sc F.~J.~Ahlers}, 
Semicond. Sci. Technol. {\bf 14}, L41 (1999).



\bibitem{Andy}
{\sc A.~S.~G.~Thornton, T.~Ihn, P.~C.~Main, L.~Eaves}, and {\sc M.~Henini}, 
Appl.\ Phys.\ Lett.\ {\bf 73}, 354 (1998).


\bibitem{wir}
{\sc I.~Hapke-Wurst, U.~Zeitler, H.~Frahm, A.~G.~M.~Jansen, R.~J.~Haug}, 
and {\sc K.~Pierz}, accepted for publication in Phys.~Rev.~B 
(see also {\sl cond-mat/0003400})

\bibitem{Kiselev}
{\sc A.~A.~Kiselev, E.~L.~Ivchenko,} and {\sc U.~R{\"o}ssler},
Phys.~Rev.~B {\bf 58}, 16353 (1998).

\bibitem{Oestreich}
{\sc M.~Oestreich, A.~P.~Heberle, W.W.~R\"uhle, R. N\"otzel,} 
and {\sc K.~Ploog},  
Europhys.~Lett.~{\bf 31}, 399 (1995).

\bibitem{LeJeune}
{\sc P.~Le Jeune, D.~Robart, X.~Marie, T.~Amand, M.~Brousseau, J.~Barrau,
V.~Kalevich,} and {\sc D.~Rodichev},
Semicond.~Sci.~Technol.~{\bf 12}, 380 (1997).

\end{references}
\end{document}